\DeclareUnicodeCharacter{03F5}{$\varepsilon$}
\documentclass[aps,prl,twocolumn,floatfix,superscriptaddress]{revtex4-2}
\usepackage{amssymb,amsmath,amstext}                
\usepackage{graphicx}                                               
\usepackage{epstopdf}                                               
\usepackage{color} 
\usepackage[dvipsnames]{xcolor}
\usepackage{bm}                                                        
\usepackage{verbatim}
\usepackage[toc,page,titletoc]{appendix}
\usepackage[utf8]{inputenc} 
\usepackage{bbold}
\usepackage{bbm}
\usepackage{ulem}
\usepackage{braket}
\usepackage[utf8]{inputenc}
\usepackage[T1]{fontenc}
\usepackage{latexsym}
\usepackage[colorlinks=true,citecolor=blue,linkcolor=magenta]{hyperref}
\usepackage[capitalise,nameinlink]{cleveref}
\crefname{supp}{Supplement}{Supplements}
\usepackage{tikz}
\usepackage{etoolbox}  

\newcommand{\mycirc}[1]{%
  \tikz[baseline=(X.base)] \node[draw, circle, inner sep=2pt] (X) {#1};%
}
\robustify{\mycirc}  

\DeclareMathOperator{\Tr}{Tr}

\DeclareMathOperator{\arctanh}{arctanh}

\def\be{\begin{equation}}
\def\ee{\end{equation}}
\def\bea{\begin{eqnarray}}
\def\eea{\end{eqnarray}}

\def\bi{\begin{itemize}}
\def\ei{\end{itemize}}
\def\ben{\begin{enumerate}}
\def\een{\end{enumerate}}

\begin{document}
\title {Aspects of quantum geometry in photonic time crystals}
\author{Karthik Subramaniam Eswaran }
\thanks{kseswaran.research@gmail.com}
\affiliation{Szkoła Doktorska Nauk Ścisłych i Przyrodniczych, Wydział Fizyki, Astronomii i Informatyki Stosowanej, Uniwersytet Jagiello\'nski, ulica Profesora Stanisława Łojasiewicza 11, PL-30-348 Kraków, Poland}
\affiliation{Instytut Fizyki Teoretycznej, Wydział Fizyki, Astronomii i Informatyki Stosowanej, Uniwersytet Jagiello\'nski, ulica Profesora Stanisława Łojasiewicza 11, PL-30-348 Kraków, Poland}
\author{Ali Emami Kopaei}
\thanks{ali.emami.app@gmail.com}
\affiliation{Szkoła Doktorska Nauk Ścisłych i Przyrodniczych, Wydział Fizyki, Astronomii i Informatyki Stosowanej, Uniwersytet Jagiello\'nski, ulica Profesora Stanisława Łojasiewicza 11, PL-30-348 Kraków, Poland}
\affiliation{Instytut Fizyki Teoretycznej, Wydział Fizyki, Astronomii i Informatyki Stosowanej, Uniwersytet Jagiello\'nski, ulica Profesora Stanisława Łojasiewicza 11, PL-30-348 Kraków, Poland}
\author{Krzysztof Sacha}
\thanks{krzysztof.sacha@uj.edu.pl}
\affiliation{Instytut Fizyki Teoretycznej, Wydział Fizyki, Astronomii i Informatyki Stosowanej, Uniwersytet Jagiello\'nski, ulica Profesora Stanisława Łojasiewicza 11, PL-30-348 Kraków, Poland}
\affiliation{Centrum Marka Kaca, Uniwersytet Jagiello\'nski, ulica Profesora Stanisława Łojasiewicza 11, PL-30-348 Kraków, Poland}

\begin{abstract}
We develop a geometric description of quantum light in photonic time crystals on the $SU(1,1)$ coherent-state manifold. In a projective picture, the evolution of each mode appears as a M\"obius isometry on the Poincar\'e disk, where topologies of trajectories distinguish stable, unstable, and critical regimes. The geometric phase is related to the hyperbolic area enclosed by cyclic paths in the complex projective Hilbert space. This framework offers an intuitive view of stability and topology in quantum photonic time crystals.
\end{abstract}

\date{\today}

\maketitle

Recently, the concept of photonic crystal—materials characterized by a spatially periodic modulation of the dielectric permittivity—  has been extended into the temporal domain through the notion of photonic time crystals (PTCs)~\cite{zurita2009reflection,salem2015temporal,PhysRevA.93.063813,doi:10.1063/1.4928659, PhysRevB.98.085142,Sounas2017,martinez2018parametric, Chamanara2018, Park2021, Lee2021, doi:10.1126/science.abo3324,asgari2024theory_,wang2024expanding,dong2025extremely}. Instead of spatial periodicity, a PTC is realized in a homogeneous medium whose optical parameters—such as dielectric permittivity—are modulated periodically in time. This temporal modulation leads to striking effects, including momentum bandgaps~\cite{biancalana2007dynamics,galiffi2022photonics,doi:10.1126/sciadv.abo6220,doi:10.1126/sciadv.adg7541,feinberg2025plasmonic}, where momentum conservation replaces energy conservation and the system is naturally described in terms of Floquet quasienergies. PTCs thus offer an entirely new degree of freedom for tailoring wave dynamics.  
The emergent phenomena in PTCs are fundamentally different from those in spatial crystals, e.g., temporal momentum band-gaps enable selective amplification of specific wavevectors, in contrast to frequency filtering in spatial photonic crystals.

In recent years, the study of PTCs has witnessed a rapid surge of interest, with numerous works exploring their properties through both analytical and numerical approaches. However, the majority of existing studies address classical or semiclassical aspects~\cite{kopaei2024towards,Eswaran_2025,PhysRevB.111.125421,PhysRevResearch.1.033069}, and the extension of these ideas into the fully quantum domain is still at an early stage of development~\cite{sustaeta2025quantum,bae2025cavity,PhysRevResearch.6.043320,Lyubarov:25}. Application of phase space or group-theoretic methods which have been tremendously successful in quantum optics~\cite{jordan1961lie,perelomov1977generalized,zachos2005quantum,arvind1995real,de1998weyl,rundle2021overview}, provides a natural route to bridge this gap. By projecting quantum evolution onto a lower-dimensional manifold with geometric structure, one can interpret stability, criticality and decoherence as direct consequences of phase space structure, describe coherence-preserving physical operations in terms of group automorphisms, and clearly define the classical-quantum correspondence. Phase space structure can also have perceptible effects on physical states in terms of the geometric phase, which depends solely on the path traversed by the system in phase space~\cite{pancharatnam1956generalized, berry1984quantal, wilczek1984appearance, aharonov1987phase, berry1988geometric, samuel1988general, wilczek1989geometric, cohen2019geometric, cisowski2022colloquium}.

In this work, we develop a geometric formulation of the quantum dynamics in PTCs by describing their evolution on the $SU(1,1)$ coherent-state manifold. Through canonical quantization of the electromagnetic field in a temporally modulated dielectric medium, we show that the evolution operator for each wave-number can be projectively mapped to a M\"obius isometry acting on the Poincar\'e disk model of the hyperbolic plane. Within this framework, the stability of bosonic modes is directly encoded in the geometry: elliptic, hyperbolic, and parabolic trajectories correspond to stable, unstable and critical parameter regimes, respectively. The onset of instability thus manifests as a topological transition in the trajectory structure, providing a direct geometric interpretation of the dynamical phase diagram of quantum PTCs. A complementary description is provided by the entropy of entanglement between generated photon pairs. The geometric phase is related to the hyperbolic area enclosed by cyclic trajectories in the complex projective Hilbert space. This geometric perspective not only deepens the theoretical understanding of PTCs and admits an interpretation of the geometric phase, but also provides an intuitive visualization of their stability landscape and entanglement dynamics.

\begin{figure}[htbp]
    \centering
    \includegraphics[width=0.45\textwidth]{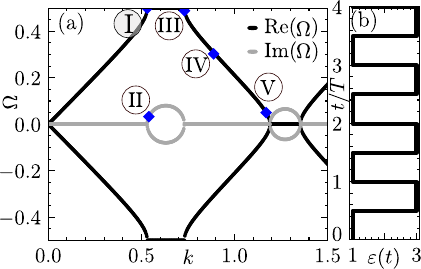} 
    \caption{(a) Band structure of the system under a temporally periodic modulation of the permittivity defined in Eq.~(\ref{epsilon}). The black dots denote the real part of the quasifrequency, while the gray dots represent its imaginary part. The blue diamonds, labeled with Roman numerals {I}–{V}, mark the points $k = \{0.531, 0.540, 0.731, 0.887, 1.169\}$, listed in the same order as the labels.
(b) Temporal profile of the step-like modulation over one period $T$, the permittivity takes the value $\varepsilon_1 = 3$ during the first half of the cycle ($0.5T$) and $\varepsilon_2 = 1$ during the second half. This periodic driving generates the band structure shown in the left panel.}
    \label{fig:1}
\end{figure}
\begin{figure*}[htbp]
    \centering
    \includegraphics[width=1\textwidth]{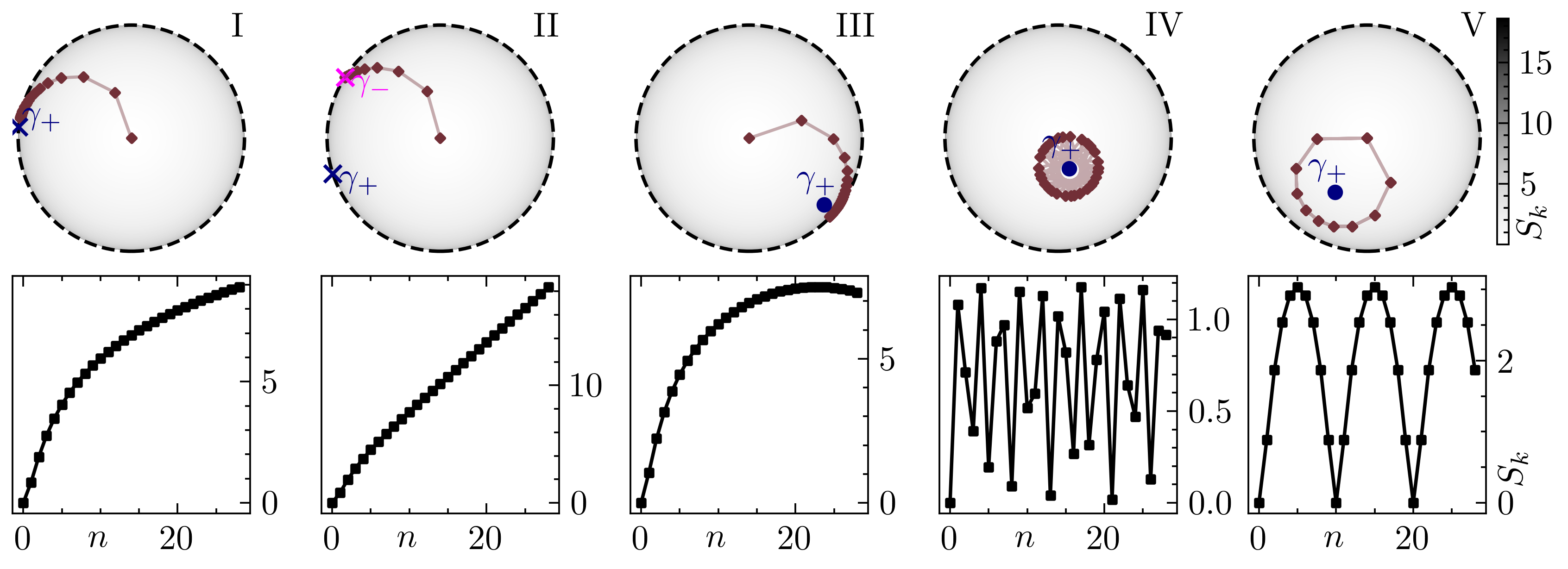} 
    \caption{Top panels: Coherent state dynamics corresponding to points I-V (see FIG.\ref{fig:1}) in the form of stroboscopic (with the period $T$) trajectories of the projective parameter $z(t)$ on the Poincar\'e disk. The points $\gamma_\pm$ correspond to fixed points of the M\"obius map $\mathcal{M}_k$. Proximity to the boundary of the disk corresponds to the number of entangled pairs. At exceptional points (e.g. I), coherent states approach the fixed points $\gamma_+=\gamma_-$ in a critically slow way. States in the gap (e.g. II) follow hyperbolic trajectories approaching one of two fixed points on the boundary. Elliptic trajectories (e.g. III-V) instead encircle a unique fixed point (for $|\,z\,|<1$). Upon approaching a critical point (III), the states begin to exhibit a critical slowing down. For incommensurate $\Omega_k$ and $\omega=2\pi/T$ (IV), trajectories display quasi-periodic behaviour. In the bottom panels, the entanglement entropy is plotted at stroboscopic times.}
    \label{fig:2}
\end{figure*}
We consider a spatially homogeneous and isotropic dielectric medium with a time-dependent relative permittivity {$\varepsilon(t)$} and constant magnetic permeability $\mu=1$. In the Coulomb gauge ($\nabla \cdot \textbf{A} = 0$) 
,the electric and magnetic fields can be expressed in terms of the vector potential $\textbf{A}$ as
$\textbf{E} = -\partial_t \textbf{A}$ and $\textbf{B} = \nabla \times \textbf{A}$. By canonically quantizing the electromagnetic field \cite{SM}, we can express the Hamiltonian of the electromagnetic field in algebraic form as{
 \begin{align}
\label{Hamil_q}
H_k(\varphi(t)) = 2\hbar kc\, e^{-\varphi} \Big[K_0\cosh{\varphi}\, 
+ \frac{K_++K_-}{2}\sinh{\varphi}\,  \Big], 
\end{align}
where $\varphi(t)=\frac{1}{2}\ln{|\varepsilon(t)|}$, and $K_0=\frac{(\hat{a}^\dagger_{\mathbf{k}}\hat{a}_{\mathbf{k}}+\hat{a}_{-\mathbf{k}}\hat{a}^\dagger_{-\mathbf{k}})}{2},\,K_+=\hat{a}^\dagger_{\mathbf{k}}\hat{a}^\dagger_{\mathbf{k}},\,K_-=K^\dagger_+$} are generators of the two-mode bosonic representation of the $SU(1,1)$ Lie group \cite{ban1993Lie-algebra} for each pair of bosonic modes $\hat{a}_{\pm \mathbf{k}}$, where $\hat{a}_{\mathbf{k}}(\hat{a}^\dagger_{\mathbf{k}})$  annihilates (creates) a photon having wavevector $\mathbf{k}$. In what follows, we assume wave propagation along the $z-$direction with wavenumber $k$. The quadratic Casimir invariant of the representation is defined as~\cite{perelomov1977generalized} 
$C_2=K^2_0-(K_+^2+K_-^2)/2= \left[  (n_k-n_{-k})^2 -1 \right] /4$, where $n_k = a^\dagger_ka_k$ is the boson number. $C_2$ commutes with every element of the algebra. As a result, unitary irreducible representations of the group can be characterized by the Minkowski momentum $P_k= \hbar k\,(n_k-n_{-k})$. Thus, the group conservation law reflects the translational invariance of the modulation, since it results in the creation of pairs of bosons having opposite momenta. Coherent states are labeled by the positive integer $m$,
\begin{equation}
\hat{K}_0\ket{\kappa,\mu}=\mu\ket{\kappa,\mu}, \quad \mu=\kappa+m,
\label{coh_state}
\end{equation}
where $\kappa$ is called the Bargmann index. The Casimir operator of this representation is given by $\hat{C}_2 = \kappa(\kappa-1)\hat{I}$. The state $\ket{\kappa,\kappa}$ of lowest weight in the discrete series representation plays a role similar to the vacuum in the formalism of generalized coherent states \cite{perelomov1986generalized} and the simplest case is the one where the initial state is the vacuum, i.e., $\kappa=1/2$ or $n_k=n_{-k}$.  For a given $\kappa$, an $SU(1,1)$ coherent state can be written in the form
\begin{equation}
\label{zdef}
\ket{z} \equiv (1-|z|^2)^{-\kappa} \sum_{m=0}^\infty \frac{(z\,\hat{K}_+)^m}{m!}\ket{\kappa,\kappa}=\mathcal{D}(\zeta)\ket{\kappa,\kappa},
\end{equation}
where, $z=\frac{\zeta}{|\zeta|} \tanh{|\zeta|}$ and ${\mathcal{D}(\zeta)=\exp{(\zeta \hat{K}_+-\zeta^*\hat{K}_-)}}$ is a displacement operator. 
For any {spatially homogeneous} modulation, the evolution operator corresponds to a M\"obius transformation acting on the unit disk{ ${D}$ {in the complex plane parametrized by $z$, i.e., ${D} =\{z\in\mathbb{C}:|z|<1\}$}. In particular,} we denote the transformation corresponding to the monodromy {for a periodic modulation $\varepsilon(t+T)=\varepsilon(t)$ having period $T$,} as $z_{n+1}=\mathcal{M}_k(z_n)$,
\begin{equation}
\label{mobiusmap}
    {\cal M}_k(z) = \frac{\alpha_k z + \beta_k}{\beta^*_k z +\alpha^*_k},
\end{equation}
where $\alpha_k,\beta_k$ parametrize the transformation and satisfy $|\alpha_k|^2-|\beta_k|^2=1$~\cite{SM}. In order to determine the evolution operators, it is convenient to work in the matrix representation of the $\mathfrak{su}(1,1)$ algebra with $\kappa=1/2$, generated by $\{\frac{1}{2}\sigma^z, \frac{i}{2}\sigma^+, \frac{i}{2}\sigma^-\}$, which act on two-component spinors. Here, $\sigma^i$ are the Pauli matrices, and $\sigma^{\pm}=(\sigma^x\pm i\sigma^y)/2$. Then, the propagator $M_k(t)\equiv U_k(t,t_0)=\mathcal{T}\exp\!\left[-i\int_{t_0}^t \! dt'\,H_k(t')\right]$, where $\mathcal{T}$ is the time-ordering operator, belongs to the matrix Lie group $SL(2,\mathbb{C})$. In the following, we denote $M_k(T)$ as $M_k$,
\begin{equation}
M_k = \left(\begin{array}{cc}
    \alpha_k & \beta_k \\
    \beta^*_k & \alpha^*_k
\end{array}\right),\quad |\alpha_k|^2-|\beta_k|^2=1.
\label{su11mono}
\end{equation}
Evaluating $U_k(t,t_0)$ corresponds to the well-known problem of disentangling time-ordered exponentials \cite{feynman51an,wei63Lie}.

We can define phase space coordinates $Z_i = \bra{z} K_i\ket{z}$ for any point $\mathbf{Z}$ lying on the $SU(1,1)$ coherent state manifold, which is the two-sheet hyperboloid $H^2=\{\mathbf{Z}\in\mathbb{R}^3\,:\,Z^2_0-Z^2_1-Z^2_2=\kappa^2\}$. Using the parametrization $z=e^{i\phi}\tanh{\nu/2}$, we have $\mathbf{Z}=(\sinh{\nu} \cos{\phi}, \sinh{\nu} \sin{\phi}, \cosh{\nu})$ for a point on $H^2$ corresponding to the coherent state parameter $z$. The same coherent state can be written as a normalized{ (with metric $\sigma^z$)} spinor,
\begin{equation}
\ket{z} = \begin{pmatrix}
    \cosh{\nu/2} \\
    ie^{-i\phi} \sinh{\nu/2},
\end{pmatrix}
\end{equation}
where $\nu\in(-\infty,\infty)$, and $\phi\in [0,\pi)$. This state can {also be written in the form} $\ket{z}=e^{i\bar{z}\sigma^-} \ket{0}/\sqrt{1-|z|^2}$. Most importantly, the Poincar\'e disk model is obtained by stereographic projection of the hyperboloid onto the unit disk in the complex plane. In this way, any state $e^{i\chi}\ket{z}$ with arbitrary phase $\chi$ is mapped to a point $z$ on the unit disk, and a propagator is mapped to the corresponding M\"obius transformation. For instance, it can be checked that $M_k\ket{z}=\exp{\left(i\arg[\alpha_k+\beta_k z^*]\right)}\ket{{\cal M}_k(z)}$. An intrinsic feature of this mapping is that since the hyperboloid is the double cover of the Poincar\'e disk, the distinct matrices $M_k,-M_k$ map to a single transformation ${\cal M}_k$, and in general a unitary rotation by angle $\theta$ of $\ket{z}$ leads to a rotation by $2\theta$ of $z$.

In the following, we consider the example of a square-wave modulation of the permittivity: for $t\in(0,T)$,
\begin{equation}
\varepsilon(t) = \begin{cases}
    \varepsilon_1 & 0<t<T_1, \\
    \varepsilon_2 & T_1<t<T.
\end{cases}
\label{epsilon}
\end{equation}
In this case, the Floquet monodromy matrix $M_k\equiv U_k(T,0)$ can be evaluated in closed form \cite{SM} using the Wei-Norman method \cite{wei63Lie}. For a differential equation with periodically varying coefficients, Floquet's theorem \cite{floquet1883equations,Shirley1965} guarantees the existence of solutions of the form
\begin{equation}
\label{floquet}
\ket{\Phi_{\Omega_k}(t)} = e^{-i\Omega_k t}\ket{\gamma_{k}(t)},
\end{equation}
where the Floquet states $\ket{\gamma_{k}(t+T)}=\ket{\gamma_{k}(t)}$ are eigenvectors of $M_k$, with {eigenvalue (i.e., Floquet multiplier) $e^{-i\Omega_kT}$}. Generally, the quasi-frequency $\Omega_k\in\mathbb{C}$, and stability of the solution requires $\mathfrak{Im}\{\Omega_k\}\le0$. The quasi-energy (i.e. quasi-frequency) spectrum, along with the modulation profile, are shown in Fig.~\ref{fig:1}.

\begin{figure}[htbp]
    \centering
    \includegraphics[width=0.45\textwidth]{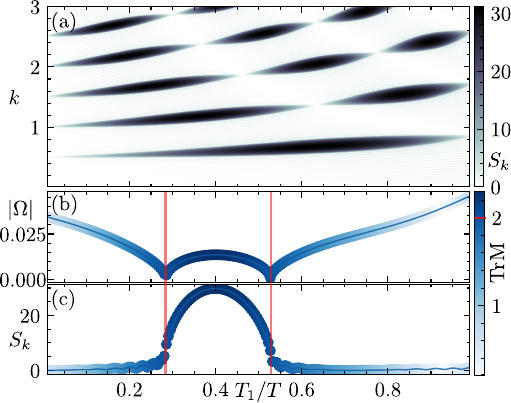} 
    \caption{(a) Phase diagram for the photonic time crystal: entanglement entropy $S_k$ can be used as an indicator of instability. (b),(c) For $k=1.22$, the drive parameter $T_1/T$ is varied across an exceptional point. Color bar corresponds to the trace of the monodromy matrix, ${\rm Tr}M_k$. The quasi-energy curve is non-analytic at exceptional points (at ratios $T_1/T$ indicated by red lines), indicating a phase transition. In all panels the bipartite entropy after $n=40$ periods, starting with the vacuum coherent state, is plotted as a function of $T_1/T$.}
    \label{fig:3}
\end{figure}

Starting from the {coherent state vaccuum}, stroboscopic trajectories of the coherent state parameter can be visualized on the Poincar\'e disk (Fig.~\ref{fig:2})~\footnote{In the interpretation of these figures, it is important to remember that the disk has a non-trivial metric: not only is the boundary at an infinite distance from the interior, but also that the shortest distance between any two points is usually not a straight line, but is rather obtained by solving the geodesic equation.}. A given $M_k$ corresponding to a transformation ${\cal M}_k$ has a geometric classification, namely it is considered to be elliptic, hyperbolic or parabolic based on its unitary equivalence class under the Iwasawa or $KAN$-decomposition \cite{iwasawa1949some, houde2024matrix}, which depends on the sign of $|\Tr{M_k}|-2$. These regimes in fact precisely correspond to whether $k$ lies on a band, in a gap or at an exceptional point, respectively. Stroboscopic trajectories are seen to take the form of closed loops when the wavenumber $k$ lies on a band. For a given coherent state $\ket{z}$, the corresponding bipartite entanglement entropy between the subsystems of $\pm k$ modes can be obtained in terms of $|z|$ as~(e.g.~\cite{li2025phase})
\begin{equation}
S_k = -\ln{\frac{1}{1-|z|^2}}-\frac{|z|^2}{1-|z|^2} \ln{|z|^2}\,.
\end{equation}
 Since the entanglement entropy depends only on $|z|$, the coherent state trajectories are closely linked to stroboscopic entanglement oscillations, regardless of phase or topology. Depending upon the distance of the state from the centre of the disk, the bipartite entanglement entropy correspondingly oscillates at a time scale proportional to $\Omega^{-1}_k$. Interestingly, when the quasi-frequency is incommensurate with the driving frequency $\omega=2\pi/T$, we observe quasi-periodic behaviour in the stroboscopic dynamics. In general, depending on the ratio $\Omega_k:\omega$, we can also observe, for example, $n-$sided polygons with a desired $n$ (see the supplementary material \cite{SM}). For a given elliptic ${\cal M}_k$, we obtain a unique fixed point $\gamma_+$ in the interior of the disk, which corresponds to the Floquet state with quasi-frequency $\pm\Omega_k$ in Eq.~(\ref{floquet}), depending on which of these has positive Bogoliubov norm, $\|\ket{w}\|_z\equiv \bra{w}{\sigma^z}\ket{w}$ (see e.g.~\cite{castin2002bose}). We choose $\ket{\gamma_k}$ as this positive norm state, which belongs to the Bargmann representation. In the hyperbolic case, a pair of fixed points, one of them attracting (e.g. \cite{milnor1990dynamics, beardon2012geometry}), appears at the boundary of the disk, related to linear divergence of the entanglement entropy, while in the parabolic case, the two fixed points coincide and are approached critically. It can be seen that the time-scale $\Omega^{-1}_k$ diverges at the transition, playing the role of a coherence time, and the entropy exhibits logarithmic divergence.

By varying the switching instant $T_1$ while maintaining a constant period [cf. (\ref{epsilon})], a given $k$ might have an elliptic, hyperbolic or parabolic monodromy (Fig.~\ref{fig:3}). Such a variation of parameters may not require re-fabrication of a given experimental setup. We can identify the points of non-analyticity of the curve $\Omega_k(T_1/T)$ as those parameters at which the transition occurs and the nature of the entanglement dynamics changes qualitatively. As $k$ increases, so does the number of stability boundaries for a variation of parameters. 

\begin{figure}[t]
    \centering
    \includegraphics[width=0.45\textwidth]{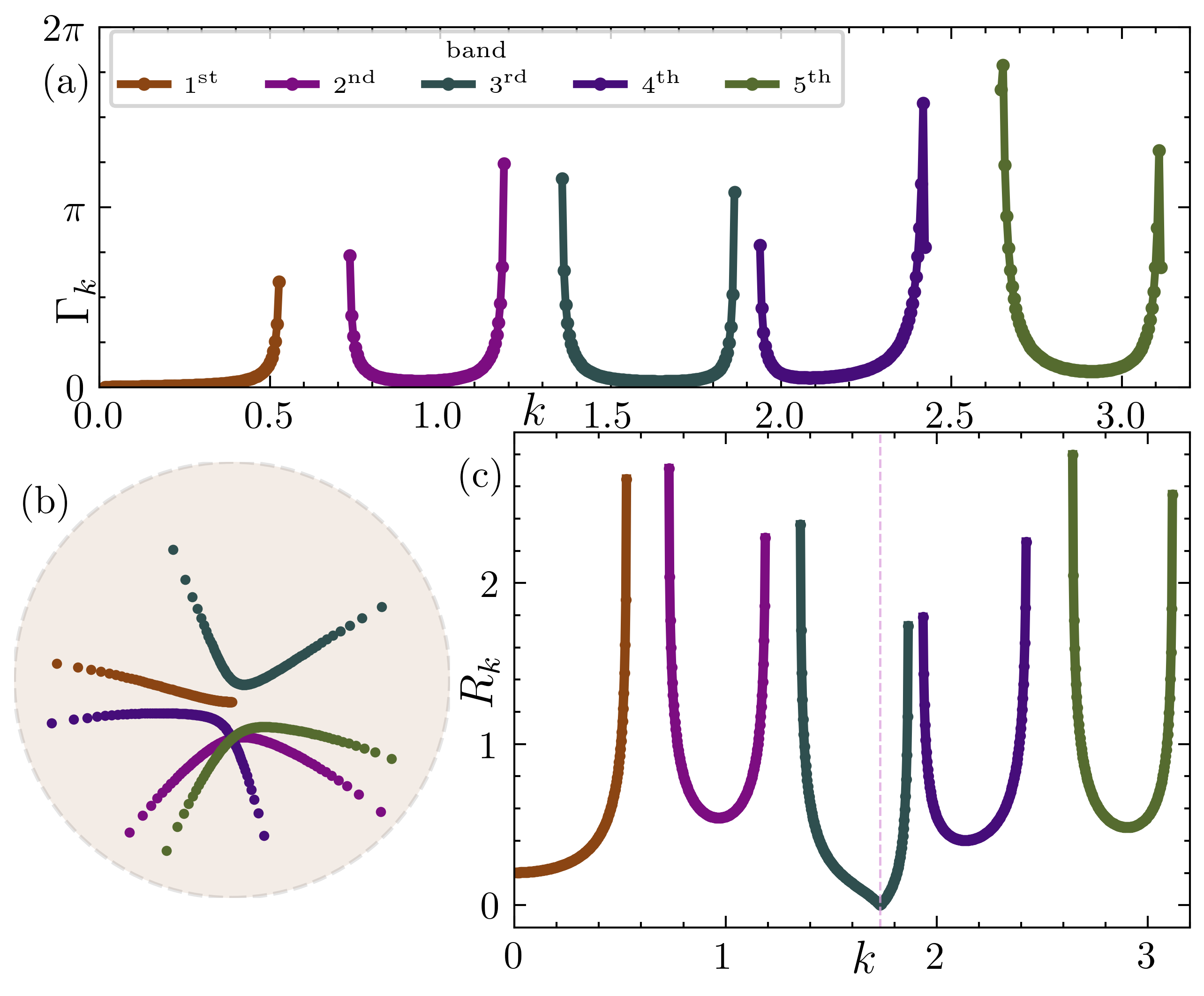} 
    \caption{(a) Geometric (Aharonov-Anandan) phase $\Gamma_k$ accumulated in the course of cyclic evolution about the stroboscopic fixed points plotted in (b). (c) Hyperbolic radii of stroboscopic trajectories starting from the vacuum state. When $R_k=0$ (at $k$ indicated by a dashed line), the vacuum is a Floquet state, representing a transparency to time refraction in the stroboscopic sense.}
    \label{fig:4}
\end{figure}

We can obtain the geometric phase as the total phase accumulated in the course of cyclic evolution in the parallel transport gauge,
{$\ket{\tilde{\gamma}_k(t)}=\text{exp}\left[i\int_0^t \text{d}t' \bra{\gamma_k(t')}{H(t')}\ket{\gamma_k(t')}\right]\ket{\gamma_k(t)}$}, as 
\begin{equation}
\Gamma_k \equiv \int_0^T\bra{\tilde{\gamma}_k(t)}i\partial_t\ket{\tilde{\gamma}_k(t)}\quad (\text{mod}\,2\pi).
\end{equation}
The geometric phase accumulated by the Floquet state $\ket{\gamma_k}$ over the course of cyclic evolution is plotted modulo $2\pi$ in Fig.~(\ref{fig:4}). The hyperbolic radius{, $R_k\equiv 2\arctanh{|\gamma_k|}$}, also presented in Fig.~(\ref{fig:4}), provides a geometric measure of how much entanglement can be generated for a given $k$ starting from the {coherent state} vacuum. Interestingly, it is possible to find $k-$values which are transparent to time refraction in the sense that the {coherent state vacuum (not to be confused with the photon vaccuum $\ket{0}_k\ket{0}_{-k}$)}  itself constitutes a Floquet state. We note that the geometric phase is seen to diverge as the stroboscopic fixed points move closer to the boundary of the disk, or equivalently, as $k$ approaches a band edge. There are several potential applications of these results. On one hand, they show that in quantum photonic time crystals, there may exist parameter regimes where the {coherent state vaccuum} is recovered after each period. Conversely, for stable solutions lying close to a band edge, the curvature of the projective phase space is reflected in a large accumulation of geometric phase per period. This implies that although the electromagnetic field for $k$-band states is stable and exhibits phase oscillations, coupling to certain $k-$modes could offer enhanced sensitivity for metrological applications, which our approach would allow one to quantify. It is also possible to envisage a scenario where we choose a drive protocol so as to completely cancel the dynamical phase in a given interval of time. For example, this would allow for the application of geometric operations on few-level quantum systems (see e.g. \cite{zhang2023geometric}), i.e., those depending purely on the photonic phase space geometry. Judging by the piecewise continuous nature of the quasi-energy curves in Fig.~(\ref{fig:3}), small variations of parameters would lead to continuous deformations of phase space trajectories as long as one does not pass through an exceptional point, which could - in principle - render these operations robust to certain types of noise.

In summary, we introduced a natural description of photonic time crystals based on coherent state representations, which provides a geometric interpretation of entanglement dynamics in the Poincar\'e disk model. Here, evolution operators take the form of M\"obius isometries, and properties of time-evolved states can be easily obtained from the analytic representation. A change in topology of coherent state trajectories was linked to stability and the wave-number band structure. Despite the rather simple structure of the Hamiltonian, the non-trivial covering of the coherent state manifold by the unit disk results in interesting topological features. Unlike states on the Poincar\'e sphere characterizing $SU(2)$, the geometric phase in photonic time crystals is not bounded, which could have several potentially interesting implications. Owing to the effectiveness of phase space methods in dealing with interactions, decoherence and information-theoretic measures, our approach might prove useful in further exploring the rich phenomenology of quantum states of light in time-varying electromagnetic media.

We acknowledge the support of the National Science
Centre, Poland, via Project No. 2021/42/A/ST2/00017.
The numerical computations in this work were supported in part by PL-Grid Infrastructure, Project No.
PLG/2025/018837.
\bibliography{reference}
\appendixpageoff
\renewcommand{\appendixtocname}{Supplementary Material}

\begin{appendices}

\setcounter{equation}{0}
\setcounter{figure}{0}
\setcounter{table}{0}

\renewcommand{\theequation}{S\arabic{equation}}
\renewcommand{\thefigure}{S\arabic{figure}}
\renewcommand{\thetable}{S\arabic{table}}

\crefalias{section}{supp}

\section{Canonical quantization of the electromagnetic field \label{model}}
We consider a spatially homogeneous and isotropic dielectric medium with a time-dependent relative permittivity $\varepsilon(t)$ and constant magnetic permeability $\mu$. Adopting the Coulomb gauge ($\nabla \cdot \textbf{A} = 0$, $\phi = 0$), the electric and magnetic fields can be expressed in terms of the vector potential $\textbf{A}$ as
$\textbf{E} = -\partial_t \textbf{A}$ and $\textbf{B} = \nabla \times \textbf{A}$. 
Within this formulation, the source-free Maxwell equations take the form
\begin{equation}
    \nabla \cdot \textbf{B} = 0, \quad 
    \nabla \times \textbf{E} = \partial_t \textbf{B}, \quad
    \nabla \cdot \textbf{D} = 0, \quad
    \nabla \times \textbf{H} = \partial_t \textbf{D}.
    \nonumber
\end{equation}
Here, the constitutive relations are given by $\textbf{D} = \varepsilon(t)\,\textbf{E}$ and $\textbf{B} = \mu\,\textbf{H}$. The Lagrangian density of the electromagnetic field can then be expressed as
\begin{equation}
    \label{lagra}
    \mathcal{L} = \tfrac{1}{2}\,\varepsilon(t)\, |\partial_t \textbf{A}|^2 - \tfrac{1}{2}\,|\nabla \times \textbf{A}|^2,
\end{equation}
where, for simplicity, we set $\mu = 1$. The corresponding Euler–Lagrange equation for the canonical coordinate $\textbf{A}(t)$ is obtained as
\begin{equation}
    \label{EL}
    \partial_t \!\left[ \varepsilon(t)\, \partial_t \textbf{A} \right] + \nabla \times \big( \nabla \times \textbf{A} \big) = 0.
\end{equation}
We now specialize to a transverse plane wave characterized by a wavevector $\textbf{k}$ and polarization along the $\textbf{z}$ direction, such that $\textbf{e}_{\textbf{k}} \perp \textbf{k}$. Within a volume $V$, the vector potential can be expressed in the mode decomposition
\begin{equation}
    \mathbf{A}(\mathbf{r}, t) = A_{\mathbf{k}}(t)\,\mathbf{e}_{\mathbf{k}}\,e^{i\mathbf{k}\cdot\mathbf{r}} + \text{h.c.},
\end{equation}
where $A_{\mathbf{k}}(t)$ denotes the time-dependent mode amplitude. Inserting this ansatz into Eq.~\eqref{EL}, the field dynamics reduce to a single-mode parametric oscillator,
\begin{align}
    \label{oscilation}
    \partial_t \left[ \varepsilon(t) \, \partial_t A_{\mathbf{k}} \right] + k^2\,c^2 A_{\mathbf{k}} = 0 
    \leftrightarrow 
    \ddot{A}_{\mathbf{k}} + \frac{\dot{\varepsilon}}{\varepsilon} &\dot{A}_{\mathbf{k}} + \omega_{\mathbf{k}}^2 A_{\mathbf{k}} = 0, \quad
    \notag \\
    \omega_{\mathbf{k}}(t) = \frac{|{\mathbf{k}}|\,c}{\sqrt{\varepsilon(t)}}.
\end{align}
where $\ \omega_{\mathbf{k}}(t)$ is the instantaneous mode frequency.
The corresponding electromagnetic energy density and Poynting theorem in a medium with time-dependent permittivity $\varepsilon(t)$ (normalized to $\varepsilon_r=1$ at the background) take the form
\begin{align}
\label{e_energyDensity}
&u = \frac{1}{2}\Big(\varepsilon(t)|\mathbf{E}|^2 + |\mathbf{B}|^2\Big), 
\notag \\
&\partial_t u + \mathbf{\nabla}\cdot(\mathbf{E}\times \mathbf{H})
= \frac{1}{2}\dot{\varepsilon}(t)\,|\mathbf{E}|^2.
\end{align}
For each mode $\mathbf{k}$, we introduce the canonical pair of variables
\begin{align}
\label{cononical}
q_{\mathbf{k}}\equiv A_{\mathbf{k}}, 
\qquad
p_{\mathbf{k}} \equiv \frac{\partial \mathcal{L}}{\partial \dot{A}_{\mathbf{k}}} = \varepsilon(t) \dot{A}_{\mathbf{k}}.
\end{align}
In terms of these canonical coordinates, the Hamiltonian of the electromagnetic field becomes
\begin{align}
\label{hamiltonian}
{\cal H}(t)
= \sum_{\mathbf{k}>0}\left[\frac{|p_{\mathbf{k}}|^2}{2\,\varepsilon(t)} + \frac{\varepsilon(t)\,\omega_{\mathbf{k}}^2(t)}{2}|q_{\mathbf{k}}|^2\right].
\end{align}
Thus, each Fourier mode of the field is equivalent to a parametric oscillator with an effective mass and frequency that vary in time.

We now proceed to quantize the field by promoting the canonical variables 
$q_{\mathbf{k}}$ and $p_{\mathbf{k}}$ to operators that satisfy the canonical commutation relation $[q_{\mathbf{k}},p_{\mathbf{k'}}] = i\hbar \delta_{\mathbf{k}\mathbf{k'}}$.
For concreteness, we take ${\mathbf{k}} = k\,\mathbf{e}_z$. Making use of the $k \leftrightarrow -k$ symmetry, we restrict to $k>0$ without loss of generality. The mode operators can then be expressed in terms of bosonic creation and annihilation operators as
\begin{align}
\label{quantiz}
&q_{\mathbf{k}} = \sqrt{\frac{\hbar}{2 k c\,V}} \,\Big(a_{\mathbf{k}} + a^\dagger_{-{\mathbf{k}}}\Big),
\notag \\ 
&p_{\mathbf{k}} = -i \sqrt{\frac{\hbar kc}{2 V}} \,\Big(a_{-\mathbf{k}}- a^\dagger_{{\mathbf{k}}}\Big),
\end{align}
with the canonical commutator $[a_{\mathbf{k}},a^{\dagger}_{\mathbf{k'}}] = \delta_{\mathbf{k}\mathbf{k'}}$.
{Here, it is useful to introduce the generators of the two-mode Bosonic representation of $SU(1,1)$},
\begin{align}
\label{su_11Gen}
&K_0 = \tfrac{1}{2}\Big(a^\dagger_{\mathbf{k}} a_{\mathbf{k}} + a^\dagger_{-{\mathbf{k}}} a_{-{\mathbf{k}}} + 1\Big), \notag \\ 
&K_1 = \tfrac{1}{2}\Big(a^\dagger_{\mathbf{k}} a^\dagger_{-{\mathbf{k}}} + a_{\mathbf{k}} a_{-{\mathbf{k}}}\Big), \\
&K_2 = \tfrac{1}{2i}\Big(a^\dagger_{\mathbf{k}} a^\dagger_{-{\mathbf{k}}} - a_{\mathbf{k}} a_{-{\mathbf{k}}}\Big), \notag
\end{align}
which obey the commutation relations $[K_0,K_1]=iK_2$, $[K_2,K_0]=iK_1$ and $[K_1,K_2]=-iK_0$. Since{Since $SU(1,1)$ is non-compact and the $K_i$ are unitary, this representation is infinite-dimensional. 
}

In terms of these generators, the Hamiltonian can be written compactly as ${\cal H}(t)=\sum_{k>0} H_k (t),$ where
 \begin{align}
H_k(t) = \hbar kc \Big[\left(1 +\varepsilon^{-1}(t)\right) K_0 
+ \left(1 - \varepsilon^{-1}(t)\right) K_1 \Big].
\end{align}

\section{Operator evolution}
First, we note that in the Pauli matrix representation, $H_k(\varphi(t))=2\hbar kc\,e^{-\varphi}(\cosh{\varphi}\,\sigma^z +i\sinh{\varphi}\,\sigma^x)$. At each instant $t,\,H^2_k(t)=4\hbar^2 k^2 c^2$, so that the instantaneous eigenprojectors $
\hbar k c \pm \frac{H_k(t)}{2}$ respectively project onto the positive and negative frequency subspaces. In this sense, symmetry under the exchange $k\to-k$ manifests as a parity symmetry in the quantum Hamiltonian. 

The Hamiltonian $H_k(t)$ can be diagonalised at each instant by a time-dependent squeezing transformation generated by $S(\varphi(t))=e^{-\varphi\,\sigma^y/2}$. The effective Hamiltonian in the resulting squeezed frame is
\begin{equation}
G_k(t)\equiv SH_kS^{-1}-iS\partial_t S^{-1}=\hbar k c e^{-\varphi}\sigma^z-\frac{i\hbar}{2}\dot{\varphi}\sigma^y.
\end{equation}
In this frame, the non-adiabatic term proportional to $\dot\varphi$ can be seen to generate entangled pairs due to two-mode squeezing generated by $i\sigma^y \cong K_2$.

We now turn to the special case of a piecewise-constant modulation, for which the time-ordered exponential for the evolution operator can be written in closed form. We consider a two-step modulation $t\in(0,T)$,
\begin{equation}
\varepsilon(t) = \begin{cases}
    \varepsilon_1 & 0<t<T_1 \\
    \varepsilon_2 & T_1<t<T,
\end{cases}
\label{espilon}
\end{equation}
so that $\dot\varphi$ takes non-zero values only at the switching instants. In this case, the group action of $M_k=U_k(T,0)$ on a vector in the original frame can be decomposed into a series of alternating two-mode squeezing and beam-splitter-like terms, as 
\be
M_k=K_{\varphi_1}U_{\omega_1T_1}K_{\varphi_2-\varphi_1}U_{\omega_2T_2}K_{-\varphi_2},
\ee
where, $T_2=T-T_1$, $\omega_i=kcT_i/\sqrt{\varepsilon_i}$, and $\varphi_i=\frac{1}{2}\ln{\left|\varepsilon_i\right|}$. Similarly for an $L-$segment piecewise continuous drive, the evolution operator takes the form $U(T,0)=\Pi_{i=1}^L K_i U_iK^{-1}_i$, generated by consecutive unitary rotations in two-mode squeezed frames.

It can easily be checked that the calculated $\alpha_k, \beta_k$ correctly reproduce the known quasi-energy dispersion relation FIG.\ref{fig:1}. The eigenvalues of $M_k$ in the elliptic or hyperbolic regimes are given by the Floquet multipliers $\rho_k=e^{-i\Omega_k T}$, satisfying
\begin{equation}
\begin{split}
\tau_k &\equiv \frac{\rho_k + \rho^{-1}_k}{2} = \cos{\left(\frac{kc\,T_1}{\sqrt{\varepsilon_1}}\right)}\cos{\left(\frac{kc\,T_2}{\sqrt{\varepsilon_2}}\right)} \\ &\quad-\frac{1}{2} \sin{\left(\frac{kc\,T_1}{\sqrt{\varepsilon_1}}\right)}\sin{\left(\frac{kc\,T_2}{\sqrt{\varepsilon_2}}\right)}\left(\sqrt{\frac{\varepsilon_1}{\varepsilon_2}}+\sqrt{\frac{\varepsilon_2}{\varepsilon_1}}\right),
\end{split}
\end{equation}
where, $\tau_k=\frac{1}{2}\text{Tr}M_k=\mathfrak{Re}\{\alpha_k\}$ and we have $\rho_k=\tau_k\pm\sqrt{\tau^2_k-1}$. We can now use this to construct the time-independant effective generator of stroboscopic dynamics (provided $M_k$ is diagonalizable), as 
\begin{equation}
\begin{split}
\label{H_eff}
H_k^\text{eff} &= \begin{cases} 
\frac{i\,\Omega_k}{\sin{(\Omega_k T)}}\left[M_k-\tau_k\right], & |\tau_k|<1;\\
\frac{\Omega_k}{\sinh{(-i\,\Omega_k T)}}\left[M_k-\tau_k\right], & |\tau_k|>1.
\end{cases}
\end{split}
\end{equation}
which is precisely the element of $\mathfrak{su}(1,1)$ corresponding to $M_k$. Note that we can combine these to write in algebraic form,
\begin{equation}
H_k^\text{eff} = \frac{\Omega_k}{\sin{(\Omega_kT)}}\left(-\mathfrak{Im}\{\alpha_k\}K_0 \,+\beta_k\,K_+\,+\bar{\beta}_k\,K_-  \right),
\end{equation}
where the hyperbolic case can be obtained by analytic continuation of $\sin{(\Omega_k T)}$ to the complex plane.

\section{Ermakov-Lewis invariant and geometric phase}
For a Hamiltonian $H(t)$, it is always possible to find a nontrivial conservation law $I(t)$, such that
\begin{equation}
\dot{I}=\partial_tI+\frac{1}{i\hbar}[I,H]=0.
\end{equation}

By explicit calculation, starting from Eq.~(\ref{hamiltonian}), we can obtain the Ermakov-Lewis invariant \cite{lewis1969exact} for non-Hermitian quadratures $q_k,p_k$ as 
\begin{equation}
\tilde{I}_k(t) = \frac{1}{2}\left[\frac{p^\dagger_k p_k}{\mu_k^2}+
\left( \mu_k q^\dagger_k + \dot{\mu}_k p_k/k^2\right)^\dagger\left( \mu_k q^\dagger_k + \dot{\mu}_k p_k/k^2\right)\right]
\end{equation}
where $\mu_k(t)$ solves the Ermakov equation,
\begin{equation}
\label{ermakov}
\ddot{\mu}_k + \frac{k^2}{\varepsilon(t)} \mu_k -\frac{1}{\mu^3_k} = 0.
\end{equation}
For constant $\varepsilon(t)=\varepsilon_c$, we can obtain an expression for $\mu_k$ by noticing that $\dot{\mu_k}$ is an integrating factor for Eq.~(\ref{ermakov}), so that a general solution for $\mu_k$ in this case reads
\begin{equation}
\mu_k(t) = \gamma_1 \left|\frac{\varepsilon_c}{k^2}\right|\left[\cosh{\delta} + \gamma_2 \sinh{\delta} \sin{\left(\frac{2\,k}{\sqrt{\varepsilon_c}}\,t+\eta\right)} \right],
\end{equation}
where $\gamma_i,\delta,\eta$ are constants of integration.

Now, defining time-dependent bosonic modes,
\begin{equation}
b_k = \frac{1}{\sqrt{2\hbar}} \left[ \frac{p_k}{\mu_k} - i(\mu_k q^\dagger_k\,+\, \dot{\mu}_k p^\dagger_k/k^2) \right],
\end{equation}
we have $[b_k,b^\dagger_{k'}]=\delta_{k,k'}$, and
\begin{equation}
\begin{split}
I(t) = \sum_k \hbar\left(b^\dagger_k\,b_k + \frac{1}{2}\right)\equiv\sum_k I_k(t),
\end{split}
\end{equation}
where we have essentially obtained a complete set of time-dependent quasinormal modes for the electromagnetic field in a time-varying medium. For completeness, we note that $\tilde{I}_k=I_k+p_kq_k-p^\dagger_kq^\dagger_k$. An eigenstate $\ket{m}_k$ of $m_k\equiv b^\dagger_k b_k$ satisfies
\begin{equation}
I_k(t)\ket{m}_k = \hbar\left(m+\frac{1}{2}\right)\ket{m}_k
\end{equation}

It is possible to write $I_k(t)$ in terms of the generators of the symmetry algebra. For the non-degenerate parametric amplifier, such an invariant can be written in terms of the 'dressed' photon number,
\begin{equation}
I_k(t) = n_k(\zeta)\,+\,n_{-k}(\zeta)\,+1 = 2K_0(\zeta(t)), 
\end{equation}
where $K_0(\zeta)=D(\zeta)K_0D^\dagger(\zeta)$ and all operators mentioned above are defined individually for a given $k$. Defining $\zeta=(\nu/2)\,\exp{(-i\phi(t))}$, by application of an identity due to Campbell, we have
\begin{equation}
\begin{split}
I_k(t) &= 2\cosh{\nu(t)}\,K_0 \\
&\quad-\,\sinh{\nu(t)}\left[e^{-i\phi(t)}\,K_+\,+\, e^{i\phi(t)}\,K_-\right]
\end{split}
\end{equation}

The eigenstates $\ket{\kappa,\mu, \zeta}$ of $I_k$ obey $I_k(t)\ket{\kappa,\mu,\zeta}=\mu\ket{\kappa,\mu,\zeta}$ and are explicitly time-dependent. Starting from the vaccuum state, they correspond to the state $\ket{\zeta} = D(\zeta)\ket{\kappa,\kappa,0}=\ket{z}$. Here, $z=\tanh{(\nu/2)}\,e^{-\phi(t)}$ is the projective Hilbert space parameter mentioned in the main text. Formally, the group of M\"obius transformations ${\cal M}_k$ under composition acting on $\{z:z\in \mathcal{D}\}$ constitutes a representation of the Lie group $SU(1,1)/U(1)$, i.e., each coherent state $\ket{z}$ corresponds to a ray in the full Hilbert space. Introducing a time-dependent phase factor,
\begin{equation}
V(t,0)\ket{z(0)}= e^{ig_{\kappa,\mu}(t)}U(t,0)\ket{z(0)},
\end{equation}
such that
\begin{equation}
\bra{z(0)}V^\dagger\dot{V}\ket{z(0)}=0.
\end{equation}
The multiplication of the state by a phase constitutes a gauge transformation, and this particular choice corresponds to a parallel transport gauge. By evolving the system, starting from the stroboscopic fixed point $\gamma_\pm$, $z(t)$ traverses a closed loop in a single time period. We thus obtain the dynamical phase $g_{\kappa,\mu}$ for an elliptic orbit,
\begin{equation}
g_{\kappa,\mu}=\int_0^{T} \bra{z(t)} H_k(t)\ket{z(t)}\,\text{d}t, 
\end{equation}
and the time-evolution operator for a closed loop,
\begin{equation}
V(T,0) = \exp{\left[ i\int_0^{T} \bra{z(t)}i\partial_t\ket{z(t)} \text{d}t\right]}.
\end{equation}
The Aharonov-Anandan phase $\Gamma_k$ is given by the eigenvalues $e^{i\Gamma_k}$ of the parallel transport gauge evolution operator for a closed loop, $V(T,0)$ traced by a Floquet state in the course of a single period. The dynamical phase can be regarded in a gauge-invariant way as the average energy of the system in a parallel transport gauge. Explicitly, with knowledge of the full micromotion trajectory, we have
\begin{equation}
\Gamma_k = 2\mu\int_0^{T} \dot{\phi} \sinh^2{\frac{\nu}{2}}\text{d}t,
\end{equation}
implying that $\Gamma_k$ is proportional to the signed area enclosed by the micromotion loop, as one might expect. 

The total phase accumulated by a system (modulo $2\pi$) in the course of evolution by a single period is determined by its Floquet quasi-energy. We can thus write for the geometric phase per period:
\begin{equation}
\Gamma_k \equiv \Omega_k T -\int_0^{T}  \bra{z(t)} H_k(t)\ket{z(t)}\,\text{d}t \quad (\text{mod}\,2\pi)
\end{equation}

Using the expression for the expectation values of the generators $K_i$ in the coherent state basis, we can readily obtain an expression for the instantaneous expectation value of $H_k(t)$ as 
\begin{equation}
\begin{split}
\bra{z(t)} &H_k(\varphi(t))\ket{z(t)} = \frac{\kappa\,\hbar kc e^{-\varphi}}{1-|z|^2}\big[ (1+|z|^2)\cosh{\varphi} \\
&\quad + 2\mathfrak{Im}\{z\}\sinh{\varphi} \big]
\end{split}
\end{equation}

for $z=z(t)$.

\section{Spinor representation}
Here we describe in detail some features of the relation between the two-component (quasi-spinor) and the holomorphic (Fock-Bargmann) representation. First, we note that a coherent state $z$ may represented as the normalized (with respect to the indefinite inner product) spinor
\begin{equation}
\begin{split}
\ket{z} &= \frac{1}{\sqrt{1-|z|^2}}\left(\begin{array}{c}
     1  \\
     i\bar{z} 
\end{array}\right) \\
&= \frac{1}{\sqrt{1-|z|^2}} e^{i\bar{z}\sigma^{-}}\left(\begin{array}{c}
     1  \\
     0 
\end{array}\right).
\end{split}
\end{equation}
Importantly, the action of a matrix $M\in SL(2,\mathbb{C}): M=\begin{pmatrix}
    \alpha & \beta \\
    \bar{\beta} & \bar{\alpha}
\end{pmatrix}$ on $\ket{z}$ reproduces the M\"obius transformation ${\cal M}:z\to \frac{\alpha z + \beta}{\bar{\beta}z+\bar{\alpha}}$ in the sense that
\begin{equation}
M\ket{z} = \exp{\big[i\arg{(\alpha + \beta \bar{z})}]}\ket{{\cal M}(z)},
\end{equation}
i.e., it maps a ray $\{\ket{z}\}$ in the projective Hilbert space onto $\{\ket{{\cal M}(z)}\}$ which corresponds to the M\"obius isometry on the unit disk. The coherent state manifold for $SU(1,1)$ is the two-sheet hyperboloid parametrized by $\nu, \phi:z=e^{i\phi}\tanh{\nu/2}$, for the upper sheet, we can write a ket upto an arbitrary phase $\chi$,
\begin{equation}
\ket{z} = e^{i\chi}\begin{pmatrix}
    \cosh{\nu/2} \\
    ie^{-i\phi} \sinh{\nu/2}.
\end{pmatrix}
\end{equation}
An important topological consequence of the map from the $\ket{z}$ to $z$ is the double covering, namely that the action of $-\mathrm{I}$ maps to the identity in $PSU(1,1)$. The relation between $PSU(1,1)\cong SO^+(2,1)$ and $SU(1,1)$ is similar to that between $SO(3)$ and $SU(2)$, except that $SU(1,1)$ is neither compact nor simply connected.
\begin{figure}[t]
    \centering
    \includegraphics[width=\columnwidth]{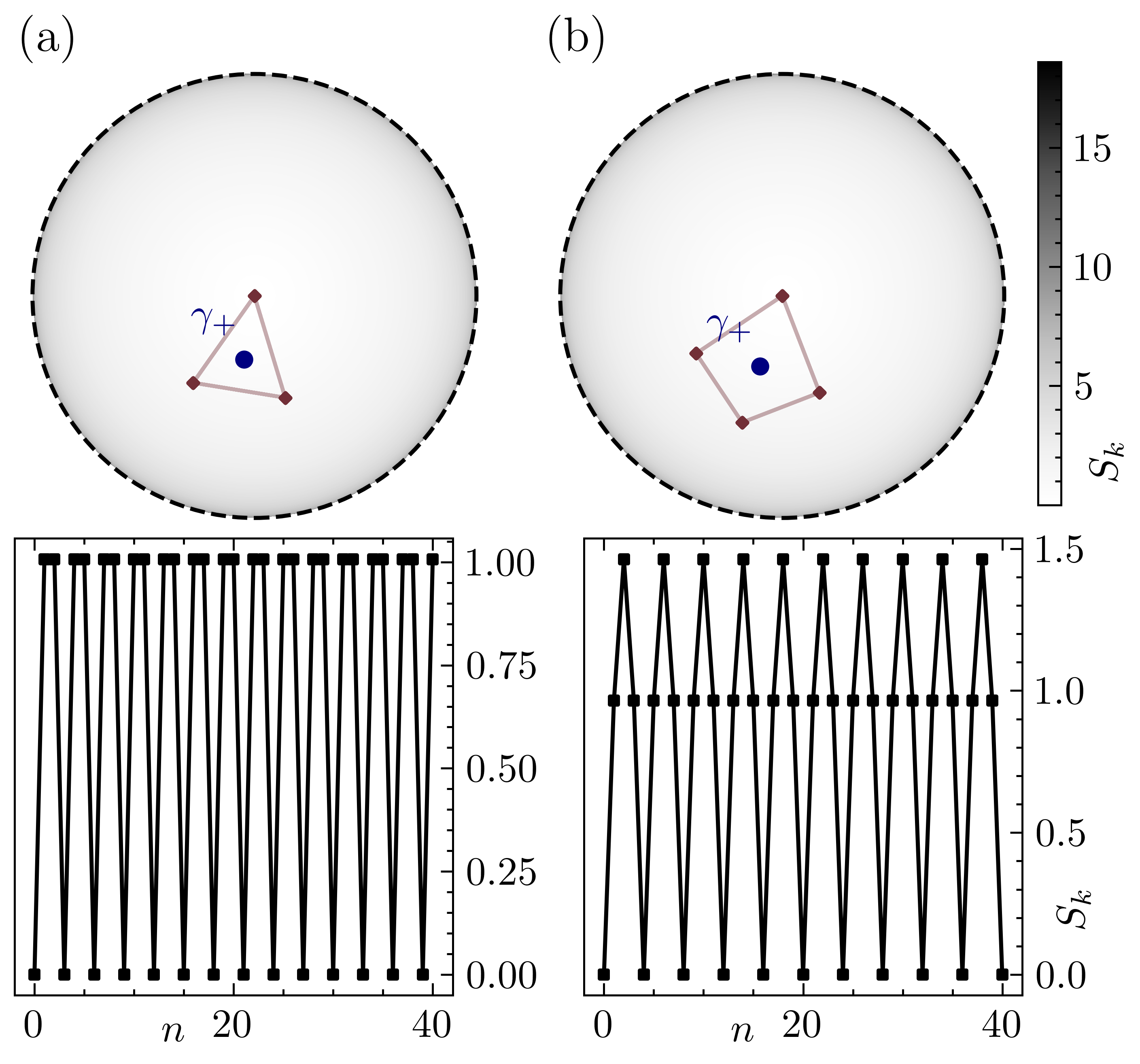}
    \caption{Stroboscopic features of coherent state evolution starting from the quasi-vaccuum state for $\Omega_k=\omega/2n$, (a) $n=3$, (b) $n=4$}
    \label{fig:app1}
\end{figure}
\section{Stroboscopic features}
Since the time period $T=2\pi/\omega$ defines a new time-scale in the system, several interesting features can emerge in stroboscopic observables depending the relation between $\omega_0(k)=kc$ and $\omega$. This was was noted, for example, in Fig.~(\ref{fig:2}) of the main text. There, the angular position about $\gamma_+$ of $z(mT)$ for $m\in\mathbb{Z}$ displpayed quasi-random behaviour with $m$. This was also reflected in the entropy series measured at stroboscopic times, and implied that quantum fluctuations of the fields are also quasi-random. On the other hand, for natural number multiplicities $\omega/2\Omega_k = n>2$, the stroboscopic trajectories form $n-$sided polygons, so that the entropies measured at stroboscopic time oscillate with a period $n$ times as long as $T$.

\end{appendices}
\end{document}